\documentclass[nofootinbib,twocolumn,showpacs,preprintnumbers,pre,aps,superscriptaddress]{revtex4-2}

\usepackage[dvipdfmx]{graphicx}
\usepackage{bm}
\usepackage{amsmath}
\usepackage{amssymb}
\usepackage{color}

\begin{document}

\title{Statistical formulation of Onsager-Machlup variational principle}

\author{Kento Yasuda}\email{yasudak@kurims.kyoto-u.ac.jp}
\author{Kenta Ishimoto}\email{ishimoto@kurims.kyoto-u.ac.jp}
\affiliation{
Research Institute for Mathematical Sciences, Kyoto University, Kyoto 606-8502, Japan}

\author{Shigeyuki Komura}\email{komura@wiucas.ac.cn}
\affiliation{Wenzhou Institute, University of Chinese Academy of Sciences, Wenzhou, Zhejiang 325001, China}
\affiliation{Oujiang Laboratory, Wenzhou, Zhejiang 325000, China}
\affiliation{Department of Chemistry, Graduate School of Science, 
Tokyo Metropolitan University, Tokyo 192-0397, Japan} 


\begin{abstract}
Onsager's variational principle (OVP) provides us with a systematic way to derive dynamical equations for various soft 
matter and active matter. 
By reformulating the Onsager-Machlup variational principle (OMVP), which is a time-global principle, we propose a new 
method to incorporate thermal fluctuations.
To demonstrate the utility of the statistical formulation of OMVP (SOMVP), we obtain the diffusion constant of a 
Brownian particle embedded in a viscous fluid by maximizing the modified Onsager-Machlup integral for the 
surrounding fluid.
We also apply our formulation to a Brownian particle in a steady shear flow, which is a typical non-equilibrium system.
Possible extensions of our formulation to internally driven active systems are also discussed.
\end{abstract}

\maketitle


\section{Introduction} 
Onsager's variational principle (OVP) is useful to obtain the governing equations of irreversible dynamics in 
soft matter~\cite{Onsager31a,Onsager31b,Doi11,DoiBook}. 
It has been applied for various problems such as colloidal suspensions~\cite{Doi11,DoiBook}, polymers~\cite{Zhou18,Doi21}, binary mixtures~\cite{Xu15}, liquid crystals~\cite{Doi11}, bilayer membranes or vesicles~\cite{Fournier15,Okamoto16,Oya18}, and liquid droplets~\cite{Man17,Hu17}.
In this principle, we construct a functional quantity called Rayleighian by summing up the dissipation 
function and the change rate of free energy.
Minimization of Rayleighian under appropriate constraints provides us with overdamped deterministic 
equations that describe the change rate of the state variables. 
The obtained equations automatically satisfy Onsager's reciprocal relations and the second law of thermodynamics.
 OVP has been applied not only for passive soft matter but also for active living systems~\cite{Zhang20,Wang21,Ackermann23}.
To take into account thermal fluctuations in OVP, noise terms are added to the deterministic equations,
and their statistical properties are determined by the fluctuation-dissipation relation~\cite{DoiBook}.

For time-evolving processes under thermal fluctuations, the Onsager-Machlup (OM) integral
can be employed to discuss the path probability~\cite{Onsager53,RiskenBook}. 
The OM integral has been used in various problems, such as structural transitions of protein 
folding~\cite{ZuckermanBook}, chemical kinetic models~\cite{Wang10}, 
and active matter~\cite{Nardini17,Wang21,Yasuda21JPSJ,Yasuda22}. 
Moreover, a path integral representation of fluctuating hydrodynamics can be formulated by 
the OM integral~\cite{Fox70,Hauge73,Bertini15,Itami15}.

Using OM integral, Doi \textit{et al.}\ proposed Onsager-Machlup variational principle (OMVP)~\cite{Doi19,Doi21}.
In contrast to OVP, which is a time-local principle, OMVP is a time-global variational principle, and it allows us to 
obtain the most probable trajectory over long time.
In particular, OMVP is useful for finding accurate solutions to the time-evolution equations when the solutions 
involve unavoidable errors due to the precision limit of numerical calculations, and it can determine long-time 
behaviors such as the steady state.
They demonstrated that OMVP can evaluate approximate solutions describing passive soft matter such as 
diffusion~\cite{Doi19}, finger flow in a square tube~\cite{Doi19}, and dye coating~\cite{Doi21}.
Recently, Wang \textit{et al.}\ used OMVP to calculate the steady state of active systems~\cite{Wang21}.
OMVP was also employed to obtain the most probable path of an active Brownian particle~\cite{Yasuda22}, 
an active Ornstein-Uhlenbeck particle~\cite{Crisanti23,Dutta23}, odd elastic systems~\cite{Yasuda21JPSJ},
and that of a chemical kinetic model~\cite{Wang10}.
However, statistical properties of stochastic trajectories and effects of thermal fluctuations, such 
as the mean square displacement of a Brownian particle, cannot be directly obtained within OMVP, 
which has been left as an important issue especially for active systems~\cite{Doi19,Doi21}.

In this paper, we propose a statistical formulation of OMVP (SOMVP) by taking into account thermal 
fluctuations of stochastic systems.
We introduce an \textit{observable} that is determined by the system variables and it allows the 
OM integral to explore trajectories deviating far from the most probable path.
We propose a modified OM integral that should be maximized to obtain the cumulant-generating function
(CGF) of the observable.
To demonstrate the usefulness of SOMVP, we calculate the diffusion constant of a Brownian 
particle embedded in a viscous fluid by optimizing the modified OM integral.
Notably, the obtained diffusion constant recovers the Stokes-Einstein relation or the fluctuation-dissipation 
relation without any additional requirements. 
To further demonstrate that SOMVP can also be used for non-equilibrium phenomena, we calculate the CGF 
of a Brownian particle subjected to a steady shear flow.
Finally, we argue possible applications of SOMVP for internally driven active matter.

\section{OVP and OMVP}
Let us consider a system that is described by the state variable $\mathbf x(\mathbf r)$ and its change rate 
$\mathbf v(\mathbf r)$.
Here, both $\mathbf x(\mathbf r)$ and $\mathbf v(\mathbf r)$ are functions of space $\mathbf r$ and 
they are independent to each other.
According to OVP, we minimize the Rayleighian $R[\mathbf v(\mathbf r),\mathbf x(\mathbf r)]$ 
with respect to $\mathbf v(\mathbf r)$ to obtain the deterministic equations for $\mathbf x(\mathbf r)$ 
and $\mathbf v(\mathbf r)$~\cite{Doi11,DoiBook}.
The Rayleighian is given by $R=\Phi+\dot F+C$, where $\Phi$ is the dissipation function, $\dot F$ is the
change rate of free energy, and $C$ represents various constraints that are included by using 
Lagrange multipliers.
The advantage of OVP is that the obtained dynamical equations automatically satisfy Onsager's reciprocal relations 
and the second law of thermodynamics.
Notice that OVP determines the instantaneous change rate $\mathbf v(\mathbf r)$ at state $\mathbf x(\mathbf r)$.

Next, we consider the time dependence of the variables, 
$\mathbf x(\mathbf r,t)$ and $\mathbf v(\mathbf r,t)$.
We introduce the OM integral that is given by the time integral of the Rayleighian~\cite{Onsager53,RiskenBook}
\begin{align}
& O[\mathbf v(\mathbf r,t),\mathbf x(\mathbf r,t)]
\nonumber \\
& =\int_0^\tau dt\, (R[\mathbf v(\mathbf r,t),\mathbf x(\mathbf r,t)]-R_\mathrm {min}
[\mathbf x(\mathbf r,t)]),
\label{OMint}
\end{align}
where $R_\mathrm {min}$ is the minimum value of $R$ in $\mathbf v$-space, i.e.,
$R_\mathrm {min}[\mathbf x(\mathbf r)]=\min_{\mathbf v(\mathbf r)} R[\mathbf v(\mathbf r),\mathbf x(\mathbf r)]$, 
and $\tau$ is the duration time.
OMVP is a global variational principle, and it states that nature chooses the path that minimizes the OM integral~\cite{Doi19}.

\section{Statistical formulation of OMVP}
In the presence of thermal fluctuations, stochastic dynamics of $\mathbf x(\mathbf r,t)$ and 
$\mathbf v(\mathbf r,t)$ involve uncertainties. 
To discuss the probability of a trajectory, Onsager and Machlup considered the path probability under 
the initial condition $\mathbf x(\mathbf r,0)=\mathbf x_0(\mathbf r)$ as given by~\cite{Onsager53,RiskenBook}
\begin{align}
P[\mathbf v(\mathbf r,t),\mathbf x(\mathbf r,t); \mathbf x_0(\mathbf r)]\sim
\exp\left(-\frac{O[\mathbf v(\mathbf r,t),\mathbf x(\mathbf r,t)]}{2k_\mathrm BT}\right),
\label{PathProb}
\end{align}
where $k_\mathrm B$ is the Boltzmann constant and $T$ is the temperature of the system.

Here, propose a statistical formulation of OMVP (SOMVP), where we introduce a 
stochastic \textit{observable} $X$ that is determined by $\mathbf v(\mathbf r,t)$ and 
$\mathbf x(\mathbf r,t)$.
Examples of $X$ are the trajectory of $\mathbf x(\mathbf r,t)$, the time average of $\mathbf v(\mathbf r,t)$ 
or $\mathbf x(\mathbf r,t)$ (see later examples), the irreversibility~\cite{Yasuda23}, and the edge current~\cite{Yashunsky22}.
Let us define the cumulant generating function (CGF) of the observable $X$ as 
\begin{align}
K_X(q)= \ln \langle\exp(qX)\rangle.
\label{DefQX}
\end{align}
In the above, the statistical average of a stochastic quantity $A$ is calculated by 
$\langle A \rangle=\int\mathcal D\mathbf v\mathcal D\mathbf x\, A P$, where $P$ is the path probability in 
Eq.~(\ref{PathProb}) and $\int\mathcal D\mathbf v\mathcal D\mathbf x$ indicates the path integral over all the 
trajectories $\mathbf v(\mathbf r,t)$ and $\mathbf x(\mathbf r,t)$.
Then, the $n$-th cumulant can be calculated from the CGF as 
\begin{align}
\left\langle X^n\right\rangle_\mathrm c=\left.\frac{d^n}{dq^n}K_X(q)\right|_{q=0}.
\end{align}

By substituting Eq.~(\ref{PathProb}) into Eq.~(\ref{DefQX}) and employing the saddle-point approximation used in the 
large deviation theory~\cite{Varadhan66,Touchette09}, the approximate CGF is given by 
\begin{align}
&K_X(q) \approx N + \max_{\mathbf v(\mathbf r,t),\mathbf x(\mathbf r,t);\mathbf x_0(\mathbf r)} 
\Omega[\mathbf v(\mathbf r,t),\mathbf x(\mathbf r,t) ]
\label{QX},\\
&\Omega[\mathbf v(\mathbf r,t),\mathbf x(\mathbf r,t) ]=qX-\frac{O[\mathbf v(\mathbf r,t),\mathbf x(\mathbf r,t) ]}{2k_\mathrm BT}+\Gamma,
\label{DifA}
\end{align}
where $N$ is the normalization factor determined by the condition $K_X(0)=0$.
Equations~(\ref{QX}) and (\ref{DifA}) are the proposed SOMVP in this paper.
In the above, $\Omega$ is the modified OM integral to be maximized with respect to both 
$\mathbf v(\mathbf r,t)$ and $\mathbf x(\mathbf r,t)$ under the given initial condition $\mathbf x_0(\mathbf r)$.
As shown in Eq.~(\ref{DifA}), the original OM integral $O$ is modified by the 
observable $X$, which allows the trajectories to deviate from the most probable path.
Such a path exploration is essential to calculate the CGF in the current framework.
In Eq.~(\ref{DifA}), a trivial connection between $\mathbf v$ and $\mathbf x$ is introduced by the term $\Gamma$
since, unlike OVP, the maximization in Eq.~(\ref{QX}) is taken with respect to both $\mathbf v$ and $\mathbf x$.
More specifically, we use $\Gamma=\int_0^\tau dt\, H_i (\dot{x}_i-v_i)$, where $\mathbf H$ is the 
Lagrange multiplier and we have used the Einstein summation convention
 (see later the case of a Brownian particle in a shear flow).

\begin{figure}[t]
\begin{center}
\includegraphics[scale=0.45]{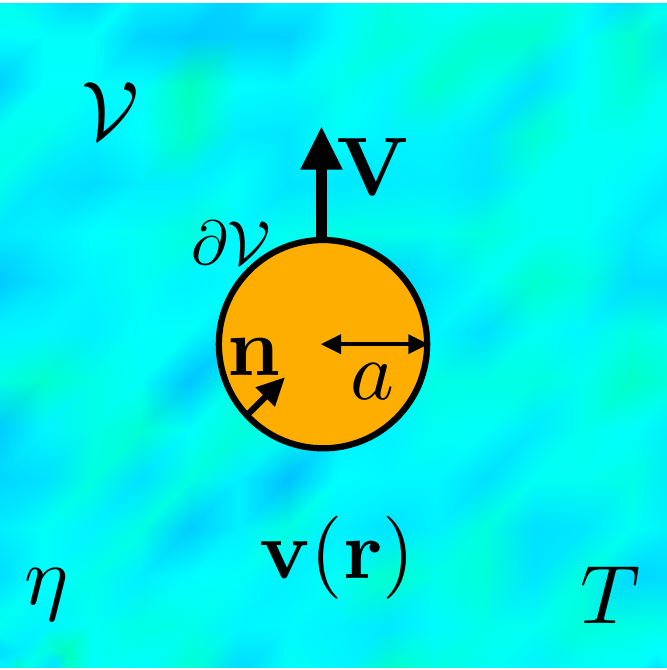}
\end{center}
\caption{
A spherical particle of radius $a$ is embedded in a three-dimensional viscous fluid with viscosity 
$\eta$ and temperature $T$.
The fluid velocity field $\mathbf v(\mathbf r)$ fluctuates due to thermal fluctuations.
The velocity of the particle $\mathbf V$ is subjected to the random forces of the surrounding fluid.
The space filled with the fluid is denoted by $\mathcal V$, whereas the particle surface is 
denoted by $\partial \mathcal V$.
}
\label{Fig:model}
\end{figure}

\section{Brownian particle in a viscous fluid}
As the simplest demonstration of SOMVP, we consider a Brownian particle embedded in 
a viscous fluid in the presence of thermal fluctuations.
In the previous works, Langevin equation of a Brownian particle was derived by solving 
the boundary problem of the fluctuating hydrodynamics~\cite{Fox70,Hauge73}.
Later, the drag coefficient of a Brownian particle was calculated by connecting the fluctuating hydrodynamics 
and Hamiltonian dynamics~\cite{Itami15}.
Although the fundamental concept of integrating the fluid degrees of freedom and casting it to the motion of 
a Brownian particle is similar to the previous approaches, we show below a more systematic derivation of the 
particle diffusion constant within SOMVP.

As shown in Fig.~\ref{Fig:model}, we consider a spherical particle of radius $a$ immersed in a fluid with 
viscosity $\eta$ and temperature $T$. 
First, the dissipation function of the fluid moving with the velocity $\mathbf v(\mathbf r)$ is given by~\cite{Doi11,DoiBook}
\begin{align}
\Phi=\frac{\eta}{4}\int_\mathcal V d^3r\, (\partial_i v_j+\partial_jv_i)^2,
\label{Phi-flu}
\end{align}
where the integral is over the three-dimensional fluid region $\mathbf r=(x,y,z)\in\mathcal V$ and  $i,j=x,y,z$.
Second, there is no free energy (potential energy) in the current problem.
Third, we employ two constraints on the fluid properties: the incompressibility condition and the stick boundary 
condition at the particle surface~\cite{HappelBrenner}.
These conditions can be expressed by the following terms  
\begin{align}
C_1&=-\int_\mathcal V d^3r\, p \partial_i v_i,\label{incomp}\\
C_2&= \int_{\partial\mathcal V} dS\, g_i(v_i-V_i),\label{boundary}
\end{align}
where $p$ and $\mathbf g$ are the Lagrange multipliers, $\mathbf V$ is the particle velocity, and 
$\partial\mathcal V$ represents the particle surface.
Notice that the particle displacement is given by $X_i(\tau)=\int_0^\tau dt \,V_i(t)$.

The Rayleighian is constructed as $R=\Phi+C_1+C_2$, and the OM integral is obtained by using Eq.~(\ref{OMint}).
In the current problem, $R_\mathrm {min}$ can be neglected since it does not depend on $\mathbf x$.
We consider the CGF of the particle displacement $\mathbf X(\tau)$ up to 
time $\tau$, i.e., $K_\mathbf X(\mathbf q)= \ln \langle\exp[q_iX_i(\tau)]\rangle$,
where $\mathbf q$ is the wave vector conjugate to $\mathbf X(\tau)$.
Since the OM integral does not depend on $\mathbf X(t)$, we obtain the modified OM integral $\Omega$
by setting $\Gamma=0$ in Eq.~(\ref{DifA}) (see Eq.~(\ref{A1})).

Following Eq.~(\ref{QX}) and maximizing $\Omega$, i.e., $\delta \Omega=0$, 
with respect to $\mathbf v$, $\mathbf V$, $p$, and $\mathbf g$, we obtain the following 
governing equations (see Appendix \ref{AppA1}):
\begin{align}
&\eta \nabla^2 v_i-\partial_i p=0,~~~\partial_i v_i=0~~~(\mathbf r\in \mathcal V),\label{StokesEq}\\
&v_i=V_i~~~(\mathbf r\in\partial \mathcal V),\label{BC}\\
&q_i-\frac{1}{2k_\mathrm BT} \int_{\partial\mathcal V} dS \,n_j\sigma_{ij}=0.
\label{q-g}
\end{align}
In Eq.~(\ref{q-g}), the stress tensor is given by $\sigma_{ij}=\eta (\partial_i v_j+\partial_jv_i)-p\delta_{ij}$,
where $p$ is the pressure.
As shown in Fig.~\ref{Fig:model}, $\mathbf n$ is the unit normal vector pointing from the fluid to the 
particle, and the fluid velocity is assumed to vanish at infinity, $\mathbf v(r\to\infty)=0$.
We first need to solve the Stokes equations in Eq.~(\ref{StokesEq}) to obtain the Stokes' law.
Then, we use Eq.~(\ref{q-g}) to calculate $\mathbf V$.

Here, we briefly summarize the solutions to the Stokes equations in the presence of a spherical 
particle~\cite{LaugaBook,GrahamBook}.
One solution is the Stokeslet $v_i^{\rm S}=f_jG_{ij}/(8\pi\eta)$ and $p^{\rm S}=f_iP_i/(8\pi)$, 
where $f_i$ are the coefficients fixed by the boundary condition, $G_{ij}=r^{-1}\delta_{ij}+r^{-3}r_ir_j$
and $P_i=2r^{-3}r_i$ with $r=|\mathbf r|$.
The other is the potential dipole solution $v_i^{\rm PD}=h_j\nabla^2G_{ij}/(8\pi\eta)$ with unknown
coefficients $h_i$ and $p^{\rm PD}=p_0$, where $p_0$ is a uniform pressure that is assumed to be zero.
Adding these two solutions, $v_i=v_i^{\rm S}+v_i^{\rm PD}$ and $p=p^{\rm S}+p^{\rm PD}$, 
we obtain the solution satisfying the boundary condition at the particle surface.
After some calculation, we obtain $f_i=6\pi\eta aV_i$ and $h_i=a^2f_i/6$ by using Eq.~(\ref{BC}).
Then, the surface integral of the stress tensor becomes
$\int_{\partial\mathcal V} dS\, n_j\sigma_{ij}=f_i$ (see Eq.~(\ref{intsigS}))~\cite{GrahamBook,LaugaBook}.
From Eq.~(\ref{q-g}), the particle velocity $\mathbf V$ can be obtained as 
\begin{align}
V_i=\frac{k_\mathrm BTq_i}{3\pi\eta a}.
\label{V-q}
\end{align}

Next, we calculate the maximized $\Omega$ by substituting the above results.
The dissipation function in Eq.~(\ref{Phi-flu}) becomes 
$\Phi=\int_{\partial\mathcal V} dS\, n_i\sigma_{ij}v_j/2=V_if_i/2$
(see Eq.~(\ref{A13}))~\cite{GrahamBook},
whereas the constraint terms disappear, $C_1=C_2=0$.
Substituting these results into $\Omega$ in Eq.~(\ref{DifA}), we obtain the CGF as 
\begin{align}
K_\mathbf X(\mathbf q)=q_i^2D\tau,
\label{StokesEinstein}
\end{align}
where the diffusion coefficient $D=k_\mathrm BT/(6\pi\eta a)$ recovers the 
Stokes-Einstein relation without any additional requirements for fluctuations.
Notice that $N$ in Eq.~(\ref{QX}) vanishes because $K_\mathbf X(0)=0$.
Then, the second-order cumulant or the mean squared displacement of the particle becomes
\begin{align}
\langle X_iX_j \rangle_\mathrm c=2D\tau\delta_{ij},
\label{XiXj}
\end{align}
recovering the required statistical property of a single Brownian particle~\cite{DoiBook}.

The above calculation demonstrates that SOMVP can be used to integrate out the stochastic 
dynamics of the surrounding fluid and to obtain the statistical properties of the macroscopic 
observable.
To further show that SOMVP also works for non-equilibrium situations, we calculate the CGF of a 
Brownian particle subjected to a steady shear flow.

\begin{figure}[t]
\begin{center}
\includegraphics[scale=0.5]{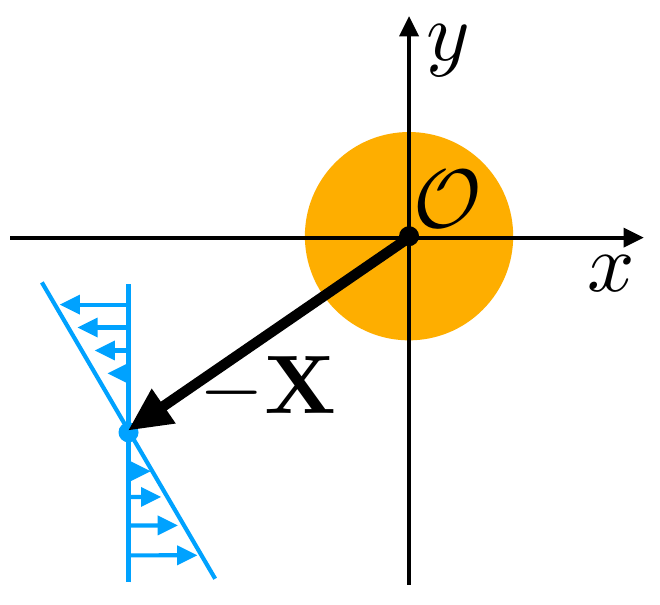}
\end{center}
\caption{
A Brownian particle in a steady shear flow.
We use the coordinate system in which the particle is fixed such that the origin of the space, $\mathcal O$, 
is fixed at the particle center.
On the other hand, the center of the shear flow (blue dot) is located at $-\mathbf X$, where $\mathbf X$ 
is the particle displacement.
}
\label{Fig:model2}
\end{figure}

\section{Brownian particle in a shear flow}
Next, we consider a Brownian particle in a steady shear flow, as shown in Fig.~\ref{Fig:model2}.
We employ a coordinate system in which the particle is fixed and the center of the shear flow is 
located at $-\mathbf X$.
Here, $\mathbf X$ is the particle displacement from the center of the shear flow.
In this problem, the flow field can be decomposed into the shear part and the remainig part 
as $\mathbf v^\mathrm{shear}+\mathbf v$, where $v_i^\mathrm{shear}=\dot\gamma_{ij} (r_j+X_j)$ 
and $\dot \gamma_{ij}$ is the shear rate tensor satisfying $\dot \gamma_{ii}=0$.

We consider the Rayleighian for the remaining part $\mathbf v$.
The dissipation function is the same as in Eq.~(\ref{Phi-flu}), and the incompressibility condition 
is also given by Eq.~(\ref{incomp}).
The boundary condition $\mathbf v^\mathrm{shear}+\mathbf v=\mathbf V$ at the particle surface 
is taken into account by the following term
\begin{align}
C_2'= \int_{\partial\mathcal V} dS\, g_i[v_i-V_i+\dot\gamma_{ij} (r_j+X_j)].\label{boundary-s}
\end{align}
Ignoring $R_\mathrm{min}$ in Eq.~(\ref{OMint}) as before, we obtain the modified OM integral $\Omega$
in Eq.~(\ref{DifA}) by using the Rayleighian $R=\Phi+C_1+C_2'$  (see Eq.~(\ref{B1})).
Unlike the previous case, $\Omega$ is a functional of both $\mathbf v$ and $\mathbf V$
as well as the particle position $\mathbf X$.
Due to the dependence on $\mathbf X$, we need an additional term 
$\Gamma=\int_0^\tau dt H_i(\dot X_i-V_i)$ in $\Omega$, ensuring $\dot {\mathbf X}=\mathbf V$.
Maximizing $\Omega$ with respect to $\mathbf v$, $\mathbf V$, $\mathbf X$, $p$, 
$\mathbf g$, and $\mathbf H$, we obtain Eq.~(\ref{StokesEq}) 
and the following set of equations (see Appendix \ref{AppB1}):
\begin{align}
&v_i=V_i-\dot\gamma_{ij} (r_j+X_j)~~~(\mathbf r\in\partial \mathcal V),\label{BC-Shear}\\
&q_i -H_i-\frac{1}{2k_\mathrm BT}\int_{\partial\mathcal V}dS\,n_j\sigma_{ij}=0,\label{q-g-shear}\\
&-\dot H_i+\dot\gamma_{ji}(q_j -H_j)=0,~~~H_i(\tau)=0.\label{Eq-H}
\end{align}

In the current problem, the solution to the Stokes equation in Eq.~(\ref{StokesEq}) can be constructed as 
$v_i=v_i^{\rm S}+v_i^{\rm PD}+v_i^{\rm SD}+v_i^{\rm PQ}$,
where $v_i^{\rm S}$ and $v_i^{\rm PD}$ are the flows due to the Stokeslet and the potential dipole solution, 
respectively, as before.
The additional terms are the Stokes dipole solution 
$v_i^{\rm SD}=\xi_{jk}\partial_kG_{ij}$, $p^{\rm SD}=\eta \xi_{ij}\partial_jP_i$ 
and the potential quadrupole solution
$v_i^{\rm PQ}=\zeta_{jk}\partial_k\nabla^2G_{ij}$, $p^{\rm PQ}=p_0$, where the coefficients 
$\xi_{ij}$ and $\zeta_{ij}$ should be fixed by the boundary condition in 
Eq.~(\ref{BC-Shear})~\cite{GrahamBook,LaugaBook}.
After some calculation (see Appendix \ref{AppB2}), all the coefficients can be determined as 
\begin{align}
&f_i=6\pi\eta a(V_i-\dot\gamma_{ij} X_j),~~~h_i=a^2f_i/6,\label{f-shear}\\
&\xi_{ij}=5a^3\dot\gamma_{ij}^{\rm S}/6+a^3\dot\gamma_{ij}^{\rm A}/2,~~~\zeta_{ij}=a^2\xi_{ij}/10,\label{xi-shear}
\end{align}
where $\dot\gamma_{ij}^{\rm S}=(\dot\gamma_{ij}+\dot\gamma_{ji})/2$ and 
$\dot\gamma_{ij}^{\rm A}=(\dot\gamma_{ij}-\dot\gamma_{ji})/2$ are the symmetric and antisymmetric
parts of the shear rate tensor, respectively.

By evaluating the surface integral of the stress tensor (see Eqs.~(\ref{intsigS}) and (\ref{B17})), Eq.~(\ref{q-g-shear}) becomes the following 
differential equation:
\begin{align}
V_i=\dot\gamma_{ij}X_j+2D(q_i -H_i),
\label{Eq-VX}
\end{align}
where $V_i=\dot X_i$.
Hereafter, we consider the simplest shear flow in the $xy$-plane,  
$\dot \gamma_{ij}=\dot\gamma\delta_{ix}\delta_{jy}$.
To find the optimum trajectory $\mathbf X(t)$, we first solve Eq.~(\ref{Eq-H}) to obtain 
$H_x=H_z=0$ and $H_y=\dot\gamma q_x(t-\tau)$, where we have used the final condition $H_i(\tau)=0$.
With these solutions, Eq.~(\ref{Eq-VX}) can be rewritten as  
$V_x=\dot\gamma Y+2Dq_x$, $V_y=-2D(H_y-q_y)$, and $V_z=2Dq_z$.
Solving these equations with the initial condition $\mathbf X_0=0$, we get the optimum trajectory as
\begin{align}
X(t)&=2D[q_xt-\dot\gamma^2 q_x(t^3/6-t^2\tau/2)+\dot\gamma q_yt^2/2],
\label{OptX}\\
Y(t)&=2D[q_yt-\dot\gamma q_x(t^2/2-t\tau)],
\label{OptY}\\
Z(t)&=2Dq_zt.
\label{OptZ}
\end{align}

The dissipation function becomes $\Phi=(V_i-\dot\gamma_{ij}X_j)f_i/2+\Phi_0$, where 
$\Phi_0=-\dot\gamma_{jk}\int_{\partial\mathcal V} dS\, n_i\sigma_{ij}r_k/2$ does not depend on $\mathbf q$
(see Eq.~(\ref{B22})).
Substituting the optimal trajectories in Eqs.~(\ref{OptX})--(\ref{OptZ}), we obtain  
\begin{align}
&\Phi-\Phi_0=2k_\mathrm BTD[q_x^2+q_z^2+\{q_y-\dot\gamma q_x(t-\tau) \}^2].
\label{OptPhi}
\end{align}
Then, the CGF in the presence of the shear flow becomes
\begin{align}
K_\mathbf X(\mathbf q)=D[q_i^2\tau+q_xq_y\dot\gamma \tau^2+q_x^2\dot\gamma^2 \tau^3/3],
\label{QXinShear}
\end{align}
where we have fixed $N$ to satisfy $K_X(0)=0$ and $\Phi_0$ has been cancelled by $N$.

Finally, the second-order cumulants are obtained from the CGF as 
\begin{align}
& \langle X^2\rangle_\mathrm c =2D\tau+2D\dot\gamma^2 \tau^3/3,
\label{XX} \\
& \langle Y^2\rangle_\mathrm c =\left\langle Z^2\right\rangle_\mathrm c=2D\tau,
\\
& \langle XY \rangle_\mathrm c =D\dot\gamma \tau^2.
\label{XY}
\end{align}
These results recover those in Refs.~\cite{Foister80,Broeck82,Schram96,Orihara11} and Eq.~(\ref{XX}) 
is different from Eq.~(\ref{XiXj}) due to the shear flow.
Notice that the correction term is proportional to $\tau^3$~\cite{Broeck82,Schram96}.
As shown in Eq.~(\ref{XY}), the shear flow also leads to the cross-correlation between $X$ and $Y$
being proportional to $\tau^2$.
From Eq.~(\ref{QXinShear}), we further find that the higher order cumulants vanish, i.e., 
$\langle\mathbf X^n\rangle_\mathrm c=0$ for $n\ge3$,
which has not been obtained by the classical approaches.
The above calculation demonstrates that SOMVP can be applied not only for systems 
in thermal equilibrium also for out-of-equilibrium systems driven by external flows.

\section{Summary and Discussion} 
In this paper, we have proposed a statistical formulation of OMVP (SOMVP) as shown in Eqs.~(\ref{QX}) 
and (\ref{DifA}).
Using this method, one can systematically obtain the cumulants of any stochastic observable.
SOMVP reproduces the established results for Brownian particles in thermal equilibrium 
[see Eqs.~(\ref{XiXj})] and in an out-of-equilibrium case such as with an applied shear flow 
[see Eqs.~(\ref{QXinShear})].
SOMVP can be applied for other stochastic systems that cannot be described by trivial governing 
equations due to complicated geometries of the problem.

The other advantage of SOMVP is that we can reuse the Rayleighian considered for different soft 
matter~\cite{DoiBook}, such as polymers~\cite{Doi21}, liquid crystals~\cite{Doi11},  
membranes surrounded by bulk fluids~\cite{Fournier15,Okamoto16}.
Moreover, various boundary conditions can be systematically taken into account in SOMVP by using 
Lagrange multipliers.
On the other hand, one limitation of the present SOMVP is that it cannot deal with active non-thermal 
fluctuations.
To include non-equilibrium fluctuations, a further extension of SOMVP is required.

Similar to the OM integral, Feynman-Kac formalism~\cite{Kac49,Wang18} and Martin-Siggia-Rose-Janssen-de Dominicis (MSRJD) formalism~\cite{MSR73,Janssen76,Dominicis78,Okamoto22} also allow us to describe stochastic dynamics via a path integral method.
Although the basic concept is common in these formulations, the problems for which these methods 
can be applied are very different.
The OM integral is particularly suitable for soft matter because the dissipation can be naturally taken 
into account in the Rayleighian to make use of the path integral method.

In this work, we have applied our formulation to an externally driven non-equilibrium system under a 
shear flow.
The method of SOMVP can also be used for internally driven active systems in which additional work $W$ is 
generated by active elements owing to the energy input or chemical reactions.
In this case, the Rayleighian is extended as $R=\Phi+\dot F+\dot W+C$ including the power of active 
work $\dot W$~\cite{Wang21}.
Following a similar procedure, one can systematically discuss statistical properties of active systems, 
such as the diffusion constant of internally driven objects.

For example, we can consider a chemically activated Janus particle for which chemical 
reactions take place on half of the particle surface~\cite{Golestanian05,Walther08}.
We use the volume fraction of the chemical product $\phi(\mathbf r)$ to express the dissipation 
function and the free energy of a binary mixture.
To describe the chemical reaction occurring on the half of the particle surface, $\partial_+\mathcal V$, 
we use the fixed-concentration condition, $\phi(\mathbf r)=1$ $(\mathbf r\in\partial_+\mathcal V)$.
Such a formulation would help to understand the enhanced self-diffusion of a catalytic enzyme 
molecule~\cite{Ghosh21}.

Another example is the observed active diffusion in living cells~\cite{Guo14} driven by active force 
dipoles~\cite{Mikhailov15,Yasuda17}.
The power due to active force dipoles can be described by 
$\dot W=cm\int_\mathcal V d^3r\,e_i(\mathbf r)e_j(\mathbf r)\partial_i v_j(\mathbf r)$, 
where $c$ is the concentration of the dipole, $m$ is the magnitude of the force dipole, and
$\mathbf e$ is a unit vector representing the direction of a dipole.
Adding $\dot W$ and the dissipation function describing the rotational motion of dipoles to the 
Rayleighian, one can discuss the Brownian motion of a tracer particle in active fluids.

In this work, we have obtained the CFG by analytically solving the obtained dynamical equations 
derived from SOMVP.
When such an approach is difficult, one can also numerically optimize the modified OM integral.
In such a situation, it may be useful to use techniques in machine learning, such as reinforcement 
learning~\cite{Cichos20,Wang22}.

\begin{acknowledgements}

We thank M.\ Doi and H.\ Wang for useful discussion.
K.Y.\ acknowledges support by a Grant-in-Aid for JSPS Fellows (Grants No.\ 22KJ1640) from the JSPS.
K.I.\ acknowledges the Japan Society for the Promotion of Science (JSPS) KAKENHI for Transformative Research Areas A (Grant No.\ 21H05309), and the Japan Science and Technology Agency (JST), FOREST (Grant No.\ JPMJFR212N). 
K.I.\ and K.Y.\ were supported in part by the Research Institute for Mathematical Sciences, an International Joint Usage/Research Center located at Kyoto University.
S.K.\ acknowledges the support by National Natural Science Foundation of China (Nos.\ 12274098 and 
12250710127) and the startup grant of Wenzhou Institute, University of Chinese Academy of Sciences 
(No.\ WIUCASQD2021041).
This work was supported by the JSPS Core-to-Core Program ``Advanced core-to-core network for the physics of self-organizing active matter" (JPJSCCA20230002).
\end{acknowledgements}

\begin{widetext}
\appendix
\section{Brownian particle in a viscous fluid}

\subsection{Derivation of Eqs.~(\ref{StokesEq}) and (\ref{q-g})}
\label{AppA1}

We show the derivation of the governing equations in Eqs.~(\ref{StokesEq}) and (\ref{q-g}).
The modified OM integral is constructed from Eqs.~(\ref{Phi-flu})--(\ref{boundary}) as 
\begin{align}
\Omega=q_i \int_0^\tau dt\, V_i-\frac{1}{2k_\mathrm BT}\int_0^\tau dt \int_\mathcal Vd^3r \,\left[\frac{\eta}{4}(\partial_i v_j+\partial_j v_i)^2-p\partial_iv_i \right]-\frac{1}{2k_\mathrm BT}\int_0^\tau dt\,\int_{\partial\mathcal V}dS\,g_i(v_i-V_i).
\label{A1}
\end{align}
According to SOMVP in Eq.~(\ref{QX}), the maximum of $\Omega$ determines CGF.
The first variation of the OM integral $\delta\Omega$ is defined by 
$\Omega[\mathbf v+\epsilon\delta\mathbf v,\mathbf V+\epsilon\delta\mathbf V]\approx \Omega[\mathbf v,\mathbf V]+\epsilon(\delta \Omega)+\epsilon^2(\delta^2 \Omega)/2$ and it is given by
\begin{align}
\delta \Omega & =q_i \int_0^\tau dt\,\delta V_i-\frac{1}{2k_\mathrm BT}\int_0^\tau dt \int_\mathcal Vd^3r \,[\eta(\partial_i v_j+\partial_j v_i)\partial_j\delta v_i-p\partial_i\delta v_i]\nonumber\\
&-\frac{1}{2k_\mathrm BT}\int_0^\tau dt \int_{\partial\mathcal V}dS \, g_i\delta v_i+\frac{1}{2k_\mathrm BT}\int_0^\tau dt\,\delta V_i\int_{\partial\mathcal V}dS\,g_i.
\end{align}
Performing integration by parts and using the incompressibility condition, $\partial_i v_i=0$, we obtain
\begin{align}
\delta \Omega&=\frac{1}{2k_\mathrm BT}\int_0^\tau dt \int_\mathcal Vd^3r \,[\eta\nabla^2 v_i-\partial_ip ]\delta v_i-\frac{1}{2k_\mathrm BT}\int_0^\tau dt \int_{\partial\mathcal V}dS \, [\eta n_j(\partial_i v_j+\partial_j v_i)-n_ip+g_i]\delta v_i\nonumber\\
&+\int_0^\tau dt\,\delta V_i\left[q_i+\frac{1}{2k_\mathrm BT}\int_{\partial\mathcal V}dS\,g_i\right].
\end{align}

For the maximization of $\Omega$ with respect to both $\mathbf v$ and $\mathbf V$, we require 
$\delta\Omega=0$.
Then, we have 
\begin{align}
&\eta\nabla^2v_i-\partial_ip=0~~~(\mathbf r\in \mathcal V),\\
&n_j\eta(\partial_i v_j+\partial_j v_i)-n_ip+g_i=0~~~(\mathbf r\in \partial \mathcal V),\\
&q_i +\frac{1}{2k_\mathrm BT}\int_{\partial\mathcal V}dS\,g_i=0.
\end{align}
Hence, we obtain the governing equations in Eqs.~(\ref{StokesEq}) and (\ref{q-g}).

\subsection{Calculation of the surface integrals}
\label{AppA2}

Next, we calculate the surface integral of the stress tensor $\sigma_{ij}$ that is necessary to determine 
$V_i$ in Eq.~(\ref{V-q}) and the dissipation function $\Phi$.
First, we write the stress tensor as follows
\begin{align}
&\sigma_{ij}=\sigma_{ij}^\mathrm S+\sigma_{ij}^\mathrm {PD},~~~\sigma_{ij}^\mathrm S=\eta(\partial_iv_j^\mathrm S+\partial_jv_i^\mathrm S)-p^\mathrm S\delta_{ij},~~~\sigma_{ij}^\mathrm{PD}=\eta(\partial_iv_j^\mathrm{PD}+\partial_jv_i^\mathrm{PD}).
\end{align}
It is known that these stress tensors are given by~\cite{GrahamBook}
\begin{align}
&\sigma_{ij}^\mathrm S=-3f_kr^{-5}r_ir_jr_k/(4\pi),~~~\sigma_{ij}^\mathrm{PD}=-3h_kr^{-5}(\delta_{ij}r_k+\delta_{ik}r_j+\delta_{jk}r_i-5r^{-2}r_ir_jr_k)/(2\pi).
\end{align}
Since the surface integral on a spherical particle is $\int_{\partial\mathcal V} dS =a^2\int_0^{2\pi} d\phi \int_0^\pi d\theta\sin\theta$, the following identities hold;
\begin{align}
&\int_{\partial\mathcal V} dS\,=4\pi a^2,~~~\int_{\partial\mathcal V} dS\, n_in_j=\frac{4\pi a^2}{3}\delta_{ij}.
\label{Id:02}
\end{align}
Using these identities and $r_i=-an_i~(\mathbf r\in \partial \mathcal V)$, we have 
\begin{align}
\int_{\partial\mathcal V} dS\, n_i\sigma_{ij}^\mathrm S=f_j,~~~\int_{\partial\mathcal V} dS\, n_i\sigma_{ij}^\mathrm{PD}=0.
\label{intsigS}
\end{align}
Hence, we obtain the well-known relation $\int_{\partial\mathcal V} dS\, n_i\sigma_{ij}=f_j$~\cite{GrahamBook,LaugaBook}.

Let us calculate the dissipation function by using the above results.
Performing integration by parts in Eq.~(\ref{Phi-flu}), we have  
\begin{align}
\Phi=\frac{\eta}{2}\left[\int_{\partial\mathcal V} dS\, n_i(\partial_i v_j+\partial_j v_i)v_j-\int_\mathcal V d^3r\, (\nabla^2 v_j) v_j\right],
\end{align}
where we have used the incompressibility condition, $\partial_iv_i=0$.
With the Stokes equation, $\eta \nabla^2 v_i-\partial_i p=0$, in Eq.~(\ref{StokesEq}), and by repeating integration by parts,
we obtain~\cite{LaugaBook}
\begin{align}
&\Phi=\frac{1}{2}\int_{\partial\mathcal V} dS\, n_i\sigma_{ij}v_j.
\end{align}
Using the boundary condition, $v_i=V_i$, in Eq.~(\ref{BC}) and the result of the surface integral of the stress tensor, we arrive at 
\begin{align}
\Phi=\frac{V_j}{2}\int_{\partial\mathcal V} dS\, n_i\sigma_{ij}=\frac{V_jf_j}{2}.
\label{A13}
\end{align}

\section{Brownian particle in a shear flow}

\subsection{Derivation of Eqs.~(\ref{StokesEq}), (\ref{q-g-shear}), and (\ref{Eq-H})}
\label{AppB1}

Here, we show the derivation of the governing equations in Eqs.~(\ref{StokesEq}), (\ref{q-g-shear}), and (\ref{Eq-H}).
The modified OM integral is constructed from Eqs.~(\ref{Phi-flu}), (\ref{incomp}), and (\ref{boundary-s}) as 
\begin{align}
\Omega & =q_i \int_0^\tau dt\, V_i-\frac{1}{2k_\mathrm BT}\int_0^\tau dt \int_\mathcal Vd^3r \,\left[\frac{\eta}{4}(\partial_i v_j+\partial_j v_i)^2-p\partial_iv_i \right]\nonumber\\
&-\frac{1}{2k_\mathrm BT}\int_0^\tau dt\,\int_{\partial\mathcal V}dS\,g_i(v_i-V_i+\dot \gamma_{ij}X_j)+\int_0^\tau dt\, H_i(\dot X_i-V_i).
\label{B1}
\end{align}
Taking the variation of $\Omega $ with respect to $\mathbf v$, $\mathbf V$, and $\mathbf X$, we obtain 
\begin{align}
\delta \Omega & =q_i \int_0^\tau dt\,\delta V_i-\frac{1}{2k_\mathrm BT}\int_0^\tau dt \int_\mathcal Vd^3r \,[\eta(\partial_i v_j+\partial_j v_i)\partial_j\delta v_i-p\partial_i\delta v_i]-\frac{1}{2k_\mathrm BT}\int_0^\tau dt \int_{\partial\mathcal V}dS \, g_i\delta v_i\nonumber\\
&+\frac{1}{2k_\mathrm BT}\int_0^\tau dt\,\delta V_i\int_{\partial\mathcal V}dS\,g_i-\frac{\dot\gamma_{ij}}{2k_\mathrm BT}\int_0^\tau dt\,\delta X_j\int_{\partial\mathcal V}dS\,g_i+\int_0^\tau dt\, H_i\delta \dot X_i-\int_0^\tau dt\,H_i\delta V_i.
\end{align}
Performing integration by parts for the space and time and using the incompressibility condition, $\partial_i v_i=0$,
we have 
\begin{align}
\delta \Omega&=\frac{1}{2k_\mathrm BT}\int_0^\tau dt \int_\mathcal Vd^3r \,[\eta\nabla^2 v_i-\partial_ip ]\delta v_i-\frac{1}{2k_\mathrm BT}\int_0^\tau dt \int_{\partial\mathcal V}dS \, [n_j\eta(\partial_i v_j+\partial_j v_i)-n_ip+g_i]\delta v_i\nonumber\\
&+\int_0^\tau dt\,\delta V_i\left[q_i +\frac{1}{2k_\mathrm BT}\int_{\partial\mathcal V}dS\,g_i\right]-\frac{\dot\gamma_{ij}}{2k_\mathrm BT}\int_0^\tau dt\,\delta X_j\int_{\partial\mathcal V}dS\,g_i\nonumber\\
&+H_i(\tau)\delta X_i(\tau)-H_i(0)\delta X_i(0)-\int_0^\tau dt\,\dot H_i\delta X_i-\int_0^\tau dt\,H_i\delta V_i.
\end{align}
Notice that $\delta X_i(0)=0$ because the initial condition $\mathbf X(0)=\mathbf X_0$ is fixed.
We require $\delta \Omega=0$ for arbitrary $\delta v_i$, $\delta V_i$, and $\delta X_i$ to obtain 
\begin{align}
&\eta\nabla^2v_i-\partial_ip=0~~~(\mathbf r\in \mathcal V),\\
&n_j\eta(\partial_i v_j+\partial_j v_i)-n_ip+g_i=0~~~(\mathbf r\in \partial \mathcal V),\\
&q_i -H_i+\frac{1}{2k_\mathrm BT}\int_{\partial\mathcal V}dS\,g_i=0,\\
&-\dot H_i-\frac{\gamma_{ji}}{2k_\mathrm BT}\int_{\partial\mathcal V}dS\,g_j=0,\\
&H_i(\tau)=0,
\end{align}
where we have used $\partial_i v_i=0$. Then, we arrive at Eqs.~(\ref{StokesEq}), (\ref{q-g-shear}), and (\ref{Eq-H}).

\subsection{Solution to the Stokes equations}
\label{AppB2}

Next, we discuss the solution to the Stokes equations in Eq.~(\ref{StokesEq}) with the boundary condition in Eq.~(\ref{BC-Shear}).
As mentioned in the main text, the solution can be constructed as $v_i=v_i^\mathrm S+v_i^\mathrm{PD}+v_i^\mathrm{SD}+v_i^\mathrm{PQ}$.
Considering the symmetry of the flow field, the boundary condition in Eq.~(\ref{BC-Shear}) can be decomposed into
two parts as 
\begin{align}
v_i^\mathrm S+v_i^\mathrm P=V_i-\dot\gamma_{ij}X_j,~~~v_i^\mathrm {SD}+v_i^\mathrm {PQ}=-\dot\gamma_{ij}r_j~~~(\mathbf r\in\partial\mathcal V).
\end{align}
For $v_i^\mathrm S+v_i^\mathrm P$, we can construct a solution by using that for a static fluid with a 
shift $V_i\to V_i-\dot\gamma_{ij} X_j$ to obtain Eq.~(\ref{f-shear}).

The additional part, $v_i^\mathrm {SD}+v_i^\mathrm {PQ}$,  has the following form;
\begin{align}
v_i^\mathrm {SD}+v_i^\mathrm {PQ}=\xi_{jk}\left(-\frac{\delta_{ij}r_k}{r^3}+\frac{\delta_{ik}r_j}{r^3}+\frac{\delta_{jk}r_i}{r^3}-3\frac{r_ir_jr_k}{r^5}\right)+6\zeta_{jk}\left(-\frac{\delta_{ij}r_k}{r^5}-\frac{\delta_{ik}r_j}{r^5}-\frac{\delta_{jk}r_i}{r^5}+\frac{5r_ir_jr_k}{r^7}\right)~~~(\mathbf r\in\mathcal V).
\end{align}
By using the relation $\zeta_{ij}=a^2\xi_{ij}/10$ , the boundary condition becomes
\begin{align}
\left[-\frac{3(\xi_{ij}+\xi_{ji})}{5}-(\xi_{ij}-\xi_{ji})+\frac{2\delta_{ij}\xi_{kk}}{5}\right]\frac{r_j}{a^3}+\dot\gamma_{ij} r_j=0~~~(\mathbf r\in\partial\mathcal V).
\end{align}
Using the relation $\xi_{ii}=0$ originating from $\dot\gamma_{ii}=0$, we have 
\begin{align}
-\frac{3(\xi_{ij}+\xi_{ji})}{5}+a^3\dot\gamma_{ij}^S=0,~~~-(\xi_{ij}-\xi_{ji})+a^3\dot\gamma_{ij}^A=0.
\end{align}
Then, we can determine the coefficient $\xi_{ij}$ as  
\begin{align}
&\xi_{ij}=5a^3\dot\gamma_{ij}^S/6+a^3\dot\gamma_{ij}^A/2,
\label{Sol:xi}
\end{align}
which is Eq.~(\ref{xi-shear}).

\subsection{Calculation of the surface integrals}
\label{AppB3}

Here, we calculate the surface integral of the stress tensor 
$\sigma_{ij}=\sigma_{ij}^\mathrm{S}+\sigma_{ij}^\mathrm{PD}+\sigma_{ij}^\mathrm{SD}+\sigma_{ij}^\mathrm{PQ}$.
We further define $\sigma_{ij}^\mathrm{SD}=\eta(\partial_j v_i^\mathrm {SD}+\partial_i v_j^\mathrm {SD})-\delta_{ij}p^\mathrm {SD}$ and $\sigma_{ij}^\mathrm{PQ}=\eta(\partial_jv_i^\mathrm {PQ}+\partial_iv_j^\mathrm {PQ})$.

First, we consider $\int_{\partial\mathcal V} dS\, n_i\sigma_{ij}$.
The surface integral of $\sigma_{ij}^\mathrm{S}$ and $\sigma_{ij}^\mathrm{PD}$ are already given in Eq.~(\ref{intsigS}).
For $\sigma_{ij}^\mathrm{SD}$ and $\sigma_{ij}^\mathrm{PQ}$, we have 
\begin{align}
\sigma_{ij}^\mathrm {SD}=&-6\eta\xi_{kl}r^{-5}\left(\delta_{il}r_jr_k+\delta_{jl}r_ir_k+\delta_{kl}r_ir_j-5r^{-2}r_ir_jr_kr_l\right),\\
\sigma_{ij}^\mathrm{PQ}=&-12\eta\zeta_{kl}r^{-5}(\delta_{ij}\delta_{kl}+\delta_{ik}\delta_{jl}+\delta_{il}\delta_{jk}-5r^{-2}(\delta_{ij}r_kr_l+\delta_{ik}r_jr_l+\delta_{il}r_jr_k+\delta_{jk}r_ir_l+\delta_{jl}r_ir_k+\delta_{kl}r_ir_j)\nonumber\\
&+35r^{-4}r_ir_jr_kr_l).
\end{align}
Using the identities
\begin{align}
&\int_{\partial\mathcal V} dS\, n_i=0,~~~\int_{\partial\mathcal V} dS\, n_in_jn_k=0,
\label{Id:13}
\end{align}
we find
\begin{align}
&\int_{\partial\mathcal V} dS\, n_i\sigma_{ij}^\mathrm{SD}=0, ~~~\int_{\partial\mathcal V} dS\, n_i\sigma_{ij}^\mathrm{PQ}=0.
\label{B17}
\end{align}
Hence, we obtain $\int_{\partial\mathcal V} dS\, n_i\sigma_{ij}=f_j$ as before~\cite{GrahamBook,LaugaBook}.
The dissipation function is 
\begin{align}
&\Phi=\frac{1}{2}\int_{\partial\mathcal V} dS\, n_i\sigma_{ij}v_j=\frac{(V_j-\dot\gamma_{jk}X_k)}{2}\int_{\partial\mathcal V} dS\, n_i\sigma_{ij}+\Phi_0,
\end{align}
with $\Phi_0=-(\dot\gamma_{jk}/2)\int_{\partial\mathcal V} dS\, n_i\sigma_{ij}r_k$.
Then the surface integral becomes $\Phi=(V_i-\dot\gamma_{ij}X_j)f_i/2+\Phi_0$ leading to Eq.~(\ref{OptPhi}) with the use 
of Eqs.~(\ref{OptX})--(\ref{OptZ}).

Second, we consider $\int_{\partial\mathcal V} dS\, n_i\sigma_{ij}r_k$ appearing in $\Phi_0$.
From Eq.~(\ref{Id:13}), we find 
\begin{align}
&\int_{\partial\mathcal V} dS\, n_i\sigma_{ij}^\mathrm{S}r_k=0, ~~~\int_{\partial\mathcal V} dS\, n_i\sigma_{ij}^\mathrm{PD}r_k=0.
\end{align}
Using the fourth-order identity
\begin{align}
&\int_{\partial\mathcal V} dS\, n_in_jn_kn_l=\frac{4\pi a^2}{15}(\delta_{ij}\delta_{kl}+\delta_{ik}\delta_{jl}+\delta_{il}\delta_{jk}),
\end{align}
and Eq.~(\ref{Id:02}), we obtain
\begin{align}
\int_{\partial\mathcal V} dS\, n_i\sigma_{ij}^\mathrm{SD}r_k=-8\pi \eta\xi_{lm}\left(-\delta_{jk}\delta_{lm}+4\delta_{jl}\delta_{km}-\delta_{jm}\delta_{kl}\right)/5,~~~
\int_{\partial\mathcal V} dS\, n_i\sigma_{ij}^\mathrm{PQ}r_k=0.
\end{align}
Using Eq.~(\ref{Sol:xi}), we get 
\begin{align}
&\Phi_0=-\dot\gamma_{jk}\int_{\partial\mathcal V} dS\, n_i\sigma_{ij}^\mathrm{SD}r_k/2=4\pi \eta\left(4\xi_{jk}\dot\gamma_{jk}-\xi_{jk}\dot\gamma_{kj}\right)/5=2\pi \eta a^3\dot\gamma_{jk}\dot\gamma_{jk},
\label{B22}
\end{align}
which is independent of $\mathbf q$.
\end{widetext}


\end{document}